\documentclass[11pt]{article}

\usepackage{amsmath,amsfonts,amssymb}

 \usepackage[numbers,sort&compress]{natbib}

\usepackage{hyperref}
\hypersetup{
   colorlinks   =  true,
    linkcolor    = blue,
    citecolor    = red,
}

\newcommand\be{\begin{equation}}
\newcommand\ee{\end{equation}}
\newcommand\bea{\begin{eqnarray}}
\newcommand\eea{\end{eqnarray}}

 \def\mf {\mathfrak}

 \newcommand\dd{{\rm d}}

  \newcommand\st{{\rm st} }
 \newcommand\til{{\rm tl}} 

\DeclareMathOperator\spn{span}

\begin{document}

\
  \vskip1cm

  \begin{center}
 {\Large \bf  
$\kappa$-Galilean and $\kappa$-Carrollian noncommutative spaces of worldlines}

\end{center}

\smallskip
 
\begin{center}

{\sc Angel Ballesteros$^{1}$, Giulia Gubitosi$^{2}$, Ivan Gutierrez-Sagredo$^{3}$, \\[4pt] and Francisco J.~Herranz$^{1}$}

\medskip
{$^1$Departamento de F\'isica, Universidad de Burgos, 
09001 Burgos, Spain}

{$^2$Dipartimento  di  Fisica  Ettore  Pancini,  Universit\`a  di  Napoli  Federico  II,  and  INFN,  Sezione  di  Napoli,  Complesso  Univ.~Monte S.~Angelo, I-80126 Napoli, Italy}

{$^3$Departamento de Matem\'aticas y Computaci\'on, Universidad de Burgos, 
09001 Burgos, Spain}
\medskip

e-mail: {\href{mailto:angelb@ubu.es}{angelb@ubu.es}, \href{mailto:igsagredo@ubu.es}{igsagredo@ubu.es},
 \href{mailto:giulia.gubitosi@unina.it}{giulia.gubitosi@unina.it},     \href{mailto:fjherranz@ubu.es}{fjherranz@ubu.es}}

\end{center}

\medskip
 \medskip

\begin{abstract}
\noindent
The noncommutative spacetimes associated to the $\kappa$-Poincar\'e relativistic symmetries and their  ``non-relativistic" (Galilei) and ``ultra-relativistic" (Carroll) limits  are indistinguishable, since their coordinates satisfy the same algebra. In this work, we show that the three quantum kinematical models can be differentiated when looking at the associated spaces of time-like worldlines. Specifically, we construct the noncommutative spaces of time-like geodesics with $\kappa$-Galilei  and $\kappa$-Carroll  symmetries as contractions of the corresponding $\kappa$-Poincar\'e space and we show that these three  spaces are defined by  different algebras. In particular, the $\kappa$-Galilei space of worldlines resembles the so-called Euclidean Snyder model, while the $\kappa$-Carroll space turns out to be commutative. Furthermore, we identify the map between quantum spaces of geodesics and the corresponding noncommutative spacetimes, which requires to extend the space of geodesics by adding  the noncommutative time coordinate. 
\end{abstract}

\medskip 
\medskip


\medskip


\newpage

\tableofcontents


\section{Introduction}

Poincar\'e transformations describe the symmetries of special relativity  and are the isometries of  Minkwoski spacetime. Their main feature is a mixing of space and time directions induced by the Lorentz transformations which generate hyperbolic rotations between them.  Such mixing is lost when taking the non-relativistic (Galilean, $c\to \infty$) or ultra-relativistic (Carrollian, $c\to 0 $) limits. In fact, Galilean and Carrollian spacetimes are characterized by degenerate metrics describing, respectively, absolute time and absolute space. 

This separation between space and time is lost in the corresponding quantum spacetimes. The first such example  was provided using  the Snyder noncommutative spacetime model~\cite{Snyder1947}, where a residual noncommutativity between the spatial and time coordinates characterizes the non-relativistic and ultra-relativistic limits (see~\cite{BGH2020snyder, BGH2020snyderPOS}).
This feature is even more striking in models with $\kappa$-deformations of relativistic symmetries. 
In fact, the noncommutative spacetimes associated to the $\kappa$-Poincar\'e~\cite{LRNT1991,GKMMK1992,LNR1992fieldtheory,Maslanka1993,MR1994,Zakrzewski1994poincare}, $\kappa$-Galilei~\cite{GKMMK1992,BGGH2020kappanewtoncarroll}, and $\kappa$-Carroll~\cite{CK4d,BGGH2020kappanewtoncarroll} symmetries  share a common    Lie-algebraic structure which is  isomorphic to the $\kappa$-Minkowski space $\mathcal{M}_\kappa$~\cite{Maslanka1993}:
 \be
[\hat x^a,\hat x^0]=\frac 1{\kappa}\, \hat x^a,\qquad [\hat x^a,\hat x^b]=0\,,
\label{a1}
\ee
where hereafter Latin indices run as $a,b=1,2,3$, while Greek ones as $\mu=0,1,2,3$.

This fact raises questions about the interpretation of quantum symmetries:  on the one hand  $\kappa$-Galilei $\mathcal{G}_\kappa$ and $\kappa$-Carroll $\mathcal{C}_\kappa$ spacetimes have exactly the same form as  $\kappa$-Minkowski; on the other hand    the quantum groups under which these spaces are covariant are different, and so are their duals, the corresponding $\kappa$-deformed quantum algebras~\cite{BGGH2020kappanewtoncarroll}.
Moreover, this degeneracy is unsatisfactory from the point of view of phenomenology:  noncommutative spacetimes are expected to model the  properties of spacetime in the quantum gravity regime (see~\cite{Addazi:2021xuf} for a recent review). Since  $\kappa$-Minkowski,   $\kappa$-Galilei and   $\kappa$-Carroll spacetimes should represent the quantum version of (3+1)D spacetimes with nonequivalent kinematical properties, it would be natural to expect that their  descriptions are different.

In this paper we show that the key to distinguish the three kinematical models and break the degeneracy is to look at the  spaces of time-like worldlines associated to each quantum group of symmetries. These quantum spaces are invariant under the $\kappa$-deformed symmetries just like the corresponding spacetimes. However, they are homogeneous spaces with respect to a different decomposition of the quantum algebras. In particular, they are defined by 6D algebras spanned by position-  and rapidity-type quantum coordinates $(\hat y^a,\hat \eta^a)$, which are dual, respectively,  to the generators of spatial translations $P_a$  and boosts $K_a$. 

The structure of this paper is as follows. We review   the constructions of  (3+1)D spacetimes and of   6D  spaces of time-like worldlines as different homogeneous spaces of the kinematical groups of symmetries in Section \ref{s2}. 
In Section \ref{s3} we revisit the construction of the quantum space of worldlines associated to the $\kappa$-Poincar\'e symmetries. In contrast to the $\kappa$-Minkowski spacetime, this quantum space is no longer of Lie-algebraic type. This was  first presented   in~\cite{BGH2019worldlinesplb} and a first phenomenological analysis was   performed in~\cite{BGGM2021fuzzy}.  By extending  the 6D $\kappa$-Poincar\'e space of worldlines  to include the noncommutative time coordinate, we can define, for the first time,  a map between the (3+1)D  $\kappa$-Minkowski spacetime and such 7D ``extended" space of worldlines. Moreover, we obtain a representation of the worldlines coordinates on  the space of functions $\Psi(\eta^1,\eta^2,\eta^3)$. This can be interpreted as the action on a classical 3D space with momentum-type coordinates $  \eta^a$, related to the 3-velocity space. In addition, we define a representation on quantum Darboux operators.   This allows to   show  that the quantum deformation parameter $\kappa^{-1}$ plays  the same algebraic role for quantum geodesics as the Planck constant $\hbar$ does for the usual phase space of quantum mechanics. 

In section \ref{s4} we construct the  novel   6D noncommutative spaces of geodesics for $\kappa$-Galilei and $\kappa$-Carroll deformations through a Lie bialgebra contraction~\cite{BGHOS1995quasiorthogonal} of the  $\kappa$-Poincar\'e  quantum worldlines. We find that the three $\kappa$-deformed  spaces of geodesics are  different: in the $\kappa$-Poincar\'e  case the commutators among $(\hat y^a,\hat \eta^a)$  involve  expressions with hyperbolic trigonometric functions depending on the  rapidities $\hat \eta^a$. These expressions reduce to a homogeneous quadratic algebra in the  $\kappa$-Galilei case, while they turn out to be commutative  in the $\kappa$-Carroll case. When extending the noncommutative spaces of worldlines to include the quantum time coordinate, also the $\kappa$-Carroll model becomes non-trivial.  Similarly to the  $\kappa$-Poincar\'e  case, the $\kappa$-Galilei extended space of worldlines  admits
 a  differential realization  on a  3D momentum-like space   on quantum Darboux operators. Hovever,  this realization is not possible in the $\kappa$-Carroll case. Additionally,   we  also present   the corresponding map
  between the quantum extended space of geodesics and the corresponding noncommutative spacetime for both the  $\kappa$-Galilei and $\kappa$-Carroll deformations.    A final section containing several remarks and open problems closes the paper.


\section{Spacetimes versus spaces of geodesics}
\label{s2}

 Before entering into the noncommutative framework, we recall the basics on homogeneous spacetimes and spaces of (time-like) lines that can be defined as homogeneous spaces of the Poincar\'e, Galilei and Carroll Lie groups. We also provide  a description of the corresponding classical coordinates  $x^\mu$ and $(  y^a,  \eta^a)$ on these spaces, thus setting up the full classical picture.

Let us consider the  (3+1)D Poincar\'e Lie algebra  in  the usual kinematical basis $\{P_0,P_a, K_a, J_a\}$ spanned, in this order,  by the generators of time translation, space translations, boosts and rotations. The  corresponding commutation rules  are given by
\be
\begin{array}{llll}
[J_a,J_b]=\epsilon_{abc}J_c ,& \quad [J_a,P_b]=\epsilon_{abc}P_c , &\quad
[J_a,K_b]=\epsilon_{abc}K_c ,  &\quad  [J_a,P_0]=0 , \\[2pt]
\displaystyle{
  [K_a,P_0]=P_a  } , &\quad\displaystyle{[K_a,P_b]=\delta_{ab} P_0} ,    &\quad\displaystyle{[K_a,K_b]=-\epsilon_{abc} J_c} , 
 &\quad 
[P_\mu,P_\nu]=0 ,
\end{array}
\label{aa}
\ee
where  sum over repeated indices is assumed.  

 The Galilei algebra is obtained as the non-relativistic limit $c\to\infty$~\cite{Inonu:1953sp} or speed-space contraction~\cite{BLL} 
  from the Poincar\'e Lie algebra.
 We   introduce explicitly the speed of light parameter $c$  in the commutation rules  (\ref{aa})  via the map
\be
 P_a\to \frac 1 c\, P_a,\qquad   K_a\to \frac 1 c\,  K_a.
\label{ab}
\ee
 Taking the limit $c\to \infty$  we get the commutation rules defining the Galilei algebra; namely
\be
\begin{array}{llll}
[J_a,J_b]=\epsilon_{abc}J_c ,& \quad [J_a,P_b]=\epsilon_{abc}P_c , &\quad
[J_a,K_b]=\epsilon_{abc}K_c ,  &\quad  [J_a,P_0]=0 , \\[2pt]
\displaystyle{
  [K_a,P_0]=P_a  } , &\quad\displaystyle{[K_a,P_b]=0} ,    &\quad\displaystyle{[K_a,K_b]= 0 } , 
 &\quad 
[P_\mu,P_\nu]=0 .
\end{array}
\label{ac}
\ee

The Carroll algebra~\cite{LevyLeblondCarroll}  is derived as the ultra-relativistic limit $c\to0$ or speed-time contraction~\cite{BLL} 
of the Poincar\'e algebra. In this case the map to be applied is
\be
 P_0\to   c\, P_0,\qquad   K_a\to  c\,  K_a,
\label{ad}
\ee
and    the limit $c\to 0$ of the commutation rules  (\ref{aa})     leads to the defining relations of the  Carroll Lie algebra
\be
\begin{array}{llll}
[J_a,J_b]=\epsilon_{abc}J_c ,& \quad [J_a,P_b]=\epsilon_{abc}P_c , &\quad
[J_a,K_b]=\epsilon_{abc}K_c ,  &\quad  [J_a,P_0]=0 , \\[2pt]
\displaystyle{
  [K_a,P_0]=0  } , &\quad\displaystyle{[K_a,P_b]=\delta_{ab} P_0} ,    &\quad\displaystyle{[K_a,K_b]=0} , 
 &\quad 
[P_\mu,P_\nu]=0 .
\end{array}
\label{ae}
\ee

Let $\mf g$ be any of the three above Lie algebras and let $G$ be the corresponding Lie group.  As a vector space,  $\mathfrak{g}$ can be written, in a generic from, as the   sum of two subspaces by means of a Cartan decomposition:
\be
{\mathfrak{g}}=  \mathfrak{t} \oplus \mathfrak{h} , \qquad  [\mathfrak{h} ,\mathfrak {h} ] \subset \mathfrak{h}  .
\label{af}
\ee
Exploiting this decomposition, we can construct a generic homogenous space  as a  left coset space $G/H$, where $H$ is the Lie group of the isotropy subalgebra $\mf h$. The generators spanning $\mathfrak{h}$ leave one point on  $G/H$ invariant. This is taken to be the origin of the space and the generators spanning $\mathfrak{h}$ act as rotations around it. On the other hand, the generators belonging to $\mathfrak{t}$ do not leave the origin invariant and are thus identified with  translations on   $G/H$.  
In this paper, we consider  the following three  homogenous spaces:
\be
\begin{array}{lll}
\multicolumn{3}{l}{ \!\!\!\!\! \mbox{$\bullet$ The (3+1)D   spacetime} \ \mathcal{S}  =G/H_\st\!:}\\[3pt]
\mathfrak g= \mathfrak t_\st \oplus \mathfrak h_\st ,  \quad & \mathfrak{t}_\st = \spn \{P_0,  {P_a} \}    ,\quad  & \mathfrak h_\st = \spn\{ {K_a},  {J_a} \}  .\\[8pt]
\multicolumn{3}{l}{ \!\!\!\!\! \mbox{$\bullet$ The 6D space of (time-like) lines} \ \mathcal{W}  =G/H_\til\!:}\\[3pt]
\mathfrak g= \mathfrak t_\til \oplus \mathfrak h_\til ,  \quad & \mathfrak{t}_\til = \spn \{  {P_a},  {K_a} \}    ,\   & \mathfrak h_\til = \spn\{P_0,  {J_a} \}=\mathbb R\oplus \mathfrak{so}(3) .\\[8pt]
\multicolumn{3}{l}{ \!\!\!\!\! \mbox{$\bullet$ The 7D extended space of (time-like) lines} \  \widetilde {\mathcal{W} } =G/\widetilde{H}_\til\!:}\\[3pt]
\mathfrak g= \widetilde{\mathfrak t}_\til \oplus \widetilde{\mathfrak h}_\til ,  \quad & \widetilde{\mathfrak{t}}_\til = \spn \{ P_0, {P_a},  {K_a} \}    ,\   & \widetilde{\mathfrak h}_\til = \spn\{  {J_a} \}= \mathfrak{so}(3) .
 \end{array}
\label{ag} 
\ee
 Recall that $H_\st$ is  just the Lorentz group ${\rm SO}(3,1)$ in the Poincar\'e case, while 
it  is isomorphic to the 3D Euclidean group ${\rm ISO}(3)$  for both the Galilei and Carroll groups.

Coordinates on the spaces $\mathcal{S}$ and $ \mathcal{W} $ can be introduced via the metric structures on these spaces. 
For the spacetime $\mathcal{S}$ one can define  Cartesian coordinates $x^\mu$ dual to the generators $P_\mu$. In terms of these coordinates the metric on $\mathcal{S}$ takes different forms in the Minkwoski, Galilei and Carroll cases, namely (see e.g.~\cite{BGH2020snyder,conf} and references therein):
\be
\begin{array}{ll}
\mbox{$\bullet$ Minkowski:}& {\rm d} s^2=( \dd x^0)^2-( \dd x^1)^2-  ( \dd x^2)^2-( \dd x^3)^2   .\\[6pt]
\mbox{$\bullet$ Galilei:}& {\rm d} s_{(1)}^2=( \dd x^0)^2 ,\qquad {\rm d} s_{(2)}^2= ( \dd x^1)^2+ ( \dd x^2)^2+( \dd x^3)^2 \ \ \mbox{on $x^0=$ constant}  .\\[6pt]
\mbox{$\bullet$ Carroll:}& {\rm d} s_{(1)}^2= ( \dd x^1)^2+ ( \dd x^2)^2+( \dd x^3)^2   ,\qquad  {\rm d} s_{(2)}^2=( \dd x^0)^2\ \ \mbox{on $x^a=$ constant}  .
 \end{array}
\label{ah} 
\ee
In the Galilei spacetime the ``main" metric $g^{(1)}$ is degenerate and corresponds to an ``absolute-time"  $x^0$. This generates a foliation (invariant under the Galilei  group action), whose leaves are defined at  constant times. A ``subsidiary"  3D non-degenerate Euclidean spatial metric   $g^{(2)}$ is restricted to each leaf of the foliation. In the Carroll spacetime the ``main" metric $g^{(1)}$ is also degenerate but  it determines  an ``absolute-space"  $x^a$. The  leaves of the invariant  foliation are defined at constant points in the 3-space. A 1D ``subsidiary"  time metric   $g^{(2)}$ is restricted to each leaf of the foliation.
Note that the quantum analogue of $x^\mu$ will be  just the $\hat x^\mu$ generators appearing in the noncommutative spacetimes (\ref{a1}). We recall that the dual features between these two classical spacetimes have been thoroughly studied (see~\cite{Duval2014} and references therein).

The metric structure on the space of worldlines $ \mathcal{W} $ (\ref{ag})  is less known; we refer to~\cite{HS1997phasespaces} for details. We denote by 
$(  y^a,  \eta^a)$ (dual to $P_a$ and $K_a$, respectively)  the six coordinates on this space. The $y^a$ are position-type coordinates and $\eta^a$ momentum-type ones (or velocities).
In the three cases characterized, respectively, by Poincar\'e, Galilei and Carroll symmetries, the metric is degenerate and there is an invariant foliation:
\be
\begin{array}{ll}
\mbox{$\bullet$ Poincar\'e:}&  {\rm d} s_{(1)}^2=(\cosh\eta ^2)^2 (\cosh\eta^3)^2(\dd \eta^1)^2+ (\cosh\eta^3)^2 (\dd \eta^2)^2+( \dd \eta^3 )^2 ,\\[6pt]
 &  {\rm d} s_{(2)}^2=( \dd y^1)^2+  ( \dd y^2)^2+( \dd y^3)^2 ,\ \ \mbox{on $\eta^a=$ constant}  .\\[6pt]
\mbox{$\bullet$ Galilei:}&  {\rm d} s_{(1)}^2=  (\dd \eta^1)^2+  (\dd \eta^2)^2+( \dd \eta^3 )^2 ,\\[6pt]
 &  {\rm d} s_{(2)}^2=( \dd y^1)^2+  ( \dd y^2)^2+( \dd y^3)^2 ,\ \ \mbox{on $\eta^a=$ constant}  .\\[6pt]
\mbox{$\bullet$ Carroll:}&  {\rm d} s_{(1)}^2= ( \dd y^1)^2+  ( \dd y^2)^2+( \dd y^3)^2  ,\\[6pt]
 &  {\rm d} s_{(2)}^2=  (\dd \eta^1)^2+  (\dd \eta^2)^2+( \dd \eta^3 )^2 ,\ \ \mbox{on $y^a=$ constant}  . 
  \end{array}
\label{ai} 
\ee
In the space of worldlines with Poincar\'e symmetries~\cite{BGH2019worldlinesplb}, the ``main" metric $g^{(1)}$ characterizes the 3-velocity space and  it  is a  hyperbolic Riemannian metric of negative constant curvature   equal to $-1/c^2$ (note that we here have set $c=1$).  
In this  3-velocity space the geodesic distance $\chi$, given by
 \be
\cosh\chi=\cosh\eta ^1 \cosh\eta ^2 \cosh\eta^3\, ,
 \label{aj}
 \ee
 corresponds to the relative rapidity between an observer at rest and one with a uniform motion with velocity $\eta^a$.  
In the space of worldlines with Galilei symmetries, the ``main" metric $g^{(1)}$ reduces to the usual one for the space of velocities in Newtonian mechanics. In this case, the geodesic distance is given by
 \be
\chi^2 \equiv \boldsymbol{\eta}^2 =  (  \eta^1)^2+  (  \eta^2)^2+(   \eta^3 )^2 \,,
 \label{ak}
 \ee
 and has the same interpretation as in the Poincar\'e case.
Finally, in the space of worldlines with Carroll symmetries, the role of 3-velocity space and 3-space is interchanged with respect to the Galilei case.
  
In the next sections, we construct the quantum counterparts  of $(  y^a,  \eta^a)$ alongside  with an additional  time coordinate $y^0$.
These seven quantum coordinates are the operators generating the $\kappa$-deformed  extended spaces of time-like worldlines,  such that  the limit $\kappa\to\infty$ leads to the commutative coordinates that we have just described.


\section{The extended space of   worldlines with  $\kappa$-Poincar\'e symmetries}
\label{s3}

We start by summarizing the main results given in~\cite{BGH2019worldlinesplb}. We recall the Poisson structure used to construct  the $\kappa$-deformation of Minkowski spacetime  $ \mathcal{S} \equiv  \mathcal{M}  =G/H_\st$ and of the space of time-like  worldlines $\mathcal{W}  =G/H_\til$   (\ref{ag}) as homogeneous spaces, in the case where $G$ is the Poincar\'e group. We then construct the   $\kappa$-deformed extended space of worldlines $\widetilde {\mathcal{W} }_\kappa$.


\subsection{The noncommutative space of time-like worldlines}

The faithful matrix representation  $\rho : \mathfrak g  \rightarrow \text{End}(\mathbb R ^5)$ for a generic element $X\in \mathfrak g $ of the   Poincar\'e Lie algebra  (\ref{aa})  is given by 
\begin{equation}
\rho(X)=   x^\mu \rho(P_\mu)  +  \xi^a \rho(K_a) +  \theta^a \rho(J_a) =
\left(\begin{array}{ccccc}
0&0&0&0&0\cr 
x^0 &0&\xi^1&\xi^2&\xi^3\cr 
x^1 &\xi^1&0&-\theta^3&\theta^2\cr 
x^2 &\xi^2&\theta^3&0&-\theta^1\cr 
x^3 &\xi^3&-\theta^2&\theta^1&0
\end{array}\right) \, .
\label{ba}
\end{equation}
The   exponentiation of $\rho$ leads to a 5D representation of the Poincar\'e group $G$. The ordering in the exponential is chosen in a way  that is suitable to define the cosets for each homogeneous space. In the case of Minkowski spacetime the group elements read
\be
\begin{array}{l}
 G_\mathcal{M}= \exp{\!\bigl(x^0 \rho(P_0)\bigr)} \exp{\!\bigl(x^1 \rho(P_1)\bigr)} \exp{\!\bigl(x^2 \rho(P_2)\bigr)} \exp{\!\bigl(x^3 \rho(P_3)\bigr)} \, H_\st,   \\[4pt]
 H_\st= \exp\bigl({\xi^1 \rho(K_1)}\bigr) \exp\bigl({\xi^2 \rho(K_2)}\bigr) \exp\bigl({\xi^3 \rho(K_3)}\bigr) \exp\bigl({\theta^1 \rho(J_1)}\bigr) \exp\bigl({\theta^2 \rho(J_2)} \bigr)\exp\bigl({\theta^3 \rho(J_3)}\bigr) .
 \end{array}
 \label{bb}
\ee
Hence $x^\mu$ constitute a well-defined set of coordinates on the (3+1)D Minkowski spacetime $ \mathcal{M} $. 
From $ G_\mathcal{M}$ we compute  left- and right-invariant vector fields and we introduce them within  the  Sklyanin bracket~\cite{ChariPressley1994} determined by the $\kappa$-Poincar\'e classical $r$-matrix~\cite{Maslanka1993,Zakrzewski1994poincare}:
\be
r=\frac{1}{\kappa} (K_1 \wedge P_1 + K_2 \wedge P_2 + K_3 \wedge P_3) .
\label{bc}
\ee
The  resulting Sklyanin brackets for the classical coordinates $x^\mu$ are just the Poisson version of the $\kappa$-Minkowski spacetime $ \mathcal{M}_\kappa$. The former is defined by linear brackets that can be directly quantized, thus giving rise to the commutation relations (\ref{a1}) for the quantum generators $\hat x^\mu$.

In order to obtain the Poisson structure on  $\mathcal{W}  =G/H_\til$,    we parametrize the Poincar\'e   group by exponentiating the  representation (\ref{ba}) with the following ordering of the exponentials:
\be
\begin{array}{l}
G_{\mathcal{W} } = \exp\bigl({\eta^1 \rho(K_1)}\bigr) \exp\bigl({y^1 \rho(P_1)}\bigr) \exp\bigl({\eta^2 \rho(K_2)}\bigr) \exp\bigl({y^2 \rho(P_2)} \bigr)\exp\bigl({\eta^3 \rho(K_3)}\bigr) \exp\bigl({y^3 \rho(P_3)}\bigr) \,H_{\til} ,   \\[4pt]
H_{\til} = \exp\bigl( {\phi^1 \rho(J_1)}\bigr) \exp\bigl( {\phi^2 \rho(J_2)}\bigr) \exp\bigl( {\phi^3 \rho(J_3)}\bigr) \exp\bigl( {y^0 \rho(P_0)} \bigr).
 \end{array}
 \label{bd}
\ee
Thus $( y^a, \eta^a)$ provide a set of coordinates for the 6D space of time-like geodesics  $\mathcal{W}$. We can deduce the  left- and right-invariant vector fields from $G_{\mathcal{W} }$ and obtain the Poisson brackets for $( y^a, \eta^a)$ through the  Sklyanin bracket with the same $r$-matrix (\ref{bc}). We find that the Poisson brackets between rapidities $\eta^a$ vanish and in the remaining brackets no ordering ambiguities  appear. Therefore the quantization is straightforward and leads to the $\kappa$-Poincar\'e  noncommutative space  of time-like geodesics $\mathcal{W}_\kappa$,  defined by the relations~\cite{BGH2019worldlinesplb}:
\begin{equation}
\begin{split}
\big[\hat y^1, \hat y^2\big] &= \frac{1}{\kappa} \left(  \sinh \hat \eta^1\, \hat y^2 -\frac{ \tanh \hat \eta^2}{\cosh \hat \eta^3}\, \hat y^1\right) ,\\
\big[\hat y^1,\hat y^3\big] &= \frac{1}{\kappa} \big( \sinh  \hat \eta^1\, \hat y^3  - \tanh  \hat \eta^3\,\hat y^1 \big), \\
\big[\hat y^2, \hat y^3\big] &= \frac{1}{\kappa} \big( \cosh  \hat \eta^1  \sinh  \hat \eta^2   \, \hat y^3 - \tanh  \hat \eta^3\,\hat y^2\big), \\
\big[\hat y^1, \hat \eta^1\big] &= \frac{1}{\kappa} \,\frac{ \bigl(\cosh  \hat \eta^1  \cosh  \hat \eta^2  \cosh  \hat \eta^3 -1\bigr)}{\cosh  \hat \eta^2  \cosh  \hat \eta^3 }, \\
\big[\hat y^2, \hat \eta^2\big] &= \frac{1}{\kappa}\, \frac{ \bigl(\cosh  \hat \eta^1  \cosh  \hat \eta^2  \cosh  \hat \eta^3 -1\bigr)}{\cosh  \hat \eta^3 }, \\
\big[\hat y^3, \hat \eta^3\big] &= \frac{1}{\kappa}\, \bigl(\cosh  \hat \eta^1  \cosh  \hat \eta^2  \cosh  \hat \eta^3 -1\bigr) \, , \\
\big[\hat \eta^a, \hat\eta^b\big] &= 0,\quad  \forall a,b,\qquad \big[\hat y^a,\hat \eta^b\big] = 0,\quad    a \neq b. 
\end{split}
\label{be}
\end{equation}
Note that the quantum analogue of  the geodesic distance (\ref{aj}), that is, 
\be
\cosh\hat \chi=\cosh\hat\eta ^1 \cosh\hat\eta ^2 \cosh\hat\eta^3 ,
\label{bf}
 \ee
 naturally appears within the commutators $[\hat y^a, \hat \eta^a]$.


 \subsection{The extended space of   worldlines}

The map between the classical coordinates of the two homogeneous spaces  $\mathcal{S}$ and $ \mathcal{W}$ can be obtained by comparing the explicit expressions of the two parametrizations of the Poincar\'e group,  $G_\mathcal{M}$ (\ref{bb}) and $G_{\mathcal{W} } $  (\ref{bd}). By doing so, one finds that  the  coordinates  $(\xi^a , \theta^a )$ coincide with   $(  \eta^a,  \phi^a)$, but the four spacetime coordinates $x^\mu$ depend in a nonlinear way on the seven    coordinates $(  y^\mu,   \eta^a)$ (see~\cite{BGH2019worldlinesplb} for the explicit expression). The fact that this mapping includes the coordinate $y^0$ suggests the definition of what we will call the 7D extended space of time-like geodesics $ \widetilde {\mathcal{W} } =G/\widetilde{H}_\til$  (\ref{ag}). The Poisson brackets satisfied by the coordinates on this space  can be computed by following the same procedure as for $\mathcal{S}$ and $ \mathcal{W}$. 

In particular, since $P_0$ commutes with the rotations $J_a$, the  same Poincar\'e group element (\ref{bd}) can be used consistently for $ \widetilde {\mathcal{W} }$, so $G_{\widetilde {\mathcal{W} }} \equiv G_{\mathcal{W} }$. This implies  that, after quantization, we obtain as a nonlinear subalgebra of $ \widetilde {\mathcal{W} }_\kappa$ the same commutators (\ref{be}) defining $\mathcal{W}_\kappa$. However, the quantization of the other Poisson brackets on $ \widetilde {\mathcal{W} }$, between $y^0$ and $y^a$, is affected by ordering ambiguities. Therefore,  quantization requires to fix the ordering $(\hat \eta^a)^m\,(\hat y^a)^n$. In this way, the additional commutation relations  that define the 
 $\kappa$-deformed  extended space of worldlines $ \widetilde {\mathcal{W} }_\kappa$ together with~\eqref{be}  turn out to be
 \begin{equation}
\begin{split}
&\big[\hat y^1, \hat y^0 \big] =  \frac 1{\kappa} \left( \hat y^1 -  \frac{\sinh \hat \eta^1 \tanh \hat \eta^2}{\cosh \hat \eta^3}\, \hat y^2 - \sinh \hat \eta^1 \tanh \hat \eta^3 \, \hat y^3 \right) ,\\
&\big[\hat y^2, \hat y^0 \big] = \frac 1{\kappa} \left( \hat y^2 +\frac{\sinh \hat \eta^1 \tanh \hat \eta^2}{\cosh \hat \eta^3}\,  \hat y^1 - \cosh \hat \eta^1 \sinh \hat \eta^2 \tanh \hat \eta^3 \, \hat y^3  \right) ,\\
&\big[\hat y^3, \hat y^0 \big] = \frac 1{\kappa} \left( \hat y^3 +  \sinh \hat \eta^1 \tanh \hat \eta^3\,\hat y^1 + \cosh \hat \eta^1 \sinh \hat \eta^2 \tanh \hat \eta^3 \,  \hat y^2\right) ,\\
&\big[\hat \eta^1, \hat y^0 \big] =  \frac 1{\kappa}\, \frac{\sinh \hat \eta^1}{\cosh \hat \eta^2 \cosh \hat \eta^3} ,\\
&\big[\hat \eta^2, \hat y^0 \big] =  \frac 1{\kappa}\, \frac{\cosh \hat \eta^1 \sinh \hat \eta^2}{\cosh \hat \eta^3} ,\\
&\big[\hat \eta^3, \hat y^0 \big] =  \frac 1{\kappa} \cosh \hat \eta^1 \cosh \hat \eta^2 \sinh \eta^3  .
\end{split}
\label{bg}
\end{equation}

Having defined the commutators for the full set of coordinates on the  extended space of worldlines $ \widetilde {\mathcal{W} }_\kappa$, eqs.~(\ref{be}) and (\ref{bg}),  one can recover the $\kappa$-Minkowski spacetime algebra  $\mathcal{M}_\kappa$  (\ref{a1}) through the quantum version of the classical nonlinear change of coordinates $ x^\mu =( y^\mu, \eta^a)$ given in~\cite{BGH2019worldlinesplb}:
\begin{align}
\begin{split}
&\hat x^0 = \cosh \hat \eta^1  \cosh \hat \eta^2  \cosh \hat \eta^3\, \hat y^0 + \sinh \hat \eta^1 \, \hat y^1 + \cosh \hat \eta^1  \sinh \hat \eta^2\, \hat y^2 + \cosh \hat \eta^1  \cosh \hat \eta^2  \sinh \hat \eta^3\, \hat y^3  ,\\
&\hat x^1  =\sinh \hat \eta^1 \cosh \hat \eta^2  \cosh \hat \eta^3\, \hat y^0 +  \cosh \hat \eta^1 \, \hat y^1+   \sinh \hat \eta^1  \sinh \hat \eta^2 \, \hat y^2+ \sinh \hat \eta^1 \cosh \hat \eta^2  \sinh \hat \eta^3\, \hat y^3   ,\\
&\hat  x^2  = \sinh \hat \eta^2  \cosh \hat \eta^3\, \hat y^0 +  \cosh \hat \eta^2\,\hat y^2  +   \sinh \hat \eta^2  \sinh \hat \eta^3\, \hat y^3, \\
&\hat x^3  =  \sinh \hat \eta^3 \, \hat y^0+ \cosh \hat \eta^3\,  \hat y^3 \, ,
\end{split}
\label{bh}
\end{align}
where the ordering $(\hat \eta^a)^m\,(\hat y^a)^n$ has to be preserved. Note that the linearization of~\eqref{bh} leads to  $\hat x^\mu \equiv \hat y^\mu$.


\subsection{Realizations on 3D momentum space and quantum Darboux operators}
\label{s31}

From the representation~\eqref{bd} one can see that the three coordinates $y^a$ are space translation parameters, while $\eta^a$ are   the rapidities associated to the three boost generators. After quantization,  the three quantum coordinates $\hat \eta^a$ generate a commutative subalgebra. Therefore, it makes sense to introduce a differential representation of the seven generators  $(\hat y^\mu,\hat \eta^a)$ of  $ \widetilde {\mathcal{W} }_\kappa$ with commutators   (\ref{be}) and (\ref{bg})   as operators acting  on the space of functions $\Psi(\eta^1,\eta^2,\eta^3)$ (that is, a representation on a classical 3D space with momentum-type coordinates $\eta^a$). Such differential realization reads
\begin{align}
\begin{split}
&\hat y^0  \Psi  = -\frac 1\kappa \left( \frac{\sinh    \eta^1}{\cosh    \eta^2  \cosh    \eta^3 }\,  \frac{\partial \Psi}{\partial \eta^1} + \frac{\cosh    \eta^1 \sinh    \eta^2 }{ \cosh    \eta^3 }\,\frac{\partial \Psi}{\partial \eta^2} +   \cosh    \eta^1 \cosh    \eta^2  \sinh    \eta^3   \,  \frac{\partial \Psi}{\partial \eta^3}   \right) ,\\[2pt]
&\hat y^1  \Psi  = \frac{1}{\kappa} \,\frac{ \bigl( \cosh  \chi -1\bigr)}{\cosh    \eta^2  \cosh    \eta^3 }\,   \frac{\partial \Psi}{\partial \eta^1}  ,\qquad 
 \hat y^2  \Psi  = \frac{1}{\kappa}\, \frac{ \bigl(\cosh  \chi -1\bigr)}{\cosh    \eta^3 } \,\frac{\partial \Psi}{\partial \eta^2}  ,\\[2pt]
&\hat y^3  \Psi  =  \frac{1}{\kappa}\, \bigl(\cosh  \chi-1\bigr) \frac{\partial \Psi}{\partial \eta^3}   ,\qquad 
  \hat \eta^a \Psi = \eta^a\Psi ,
\end{split}
\label{bi}
\end{align}
where $\cosh  \chi$ is given in   (\ref{aj}). As a byproduct, by introducing this result into  the relations (\ref{bh})   we  obtain the   differential realization of the $\kappa$-Minkowski  algebra (\ref{a1}) on the same 3D momentum space:
\begin{align}
\begin{split}
&\hat x^1  \Psi  =- \frac 1\kappa \left( \bigg( \frac{\cosh  \eta^1 }{\cosh    \eta^2 \cosh    \eta^3 }-1 \biggr) \frac{\partial \Psi}{\partial \eta^1} + \frac{\sinh    \eta^1\sinh    \eta^2 }{ \cosh    \eta^3 }\, \frac{\partial \Psi}{\partial \eta^2} +  {\sinh    \eta^1 \cosh    \eta^2  \sinh    \eta^3}\, \frac{\partial \Psi}{\partial \eta^3}   \right) ,\\[2pt]
&\hat x^2  \Psi  =- \frac 1\kappa \left(  {\sinh    \eta^1\tanh    \eta^2 } \, \frac{\partial \Psi}{\partial \eta^1} + \bigg( \frac{\cosh    \eta^2 }{\cosh    \eta^3  }-\cosh    \eta^1 \biggr)  \frac{\partial \Psi}{\partial \eta^2} +  {  \sinh    \eta^2  \sinh    \eta^3}\, \frac{\partial \Psi}{\partial \eta^3}   \right) ,\\[2pt]
&\hat x^3  \Psi  = -\frac 1\kappa \left(   \frac{\sinh    \eta^1\tanh    \eta^3 }{\cosh    \eta^2   }\,\frac{\partial \Psi}{\partial \eta^1} +\cosh  \eta^1\sinh  \eta^2 \tanh    \eta^3\, \frac{\partial \Psi}{\partial \eta^2} -\bigl(\cosh  \eta^1\cosh  \eta^2-\cosh  \eta^3 \bigr) \frac{\partial \Psi}{\partial \eta^3}   \right) \, ,
\end{split}
\label{bk}
\end{align}
such that the action of the time operator is just $\hat x^0  \Psi \equiv \hat y^0  \Psi $ given by  (\ref{bi}).

We can also define the $\kappa$-deformed space of worldlines ${\mathcal{W} }_\kappa$ in terms of   ``quantum Darboux operators" $(\hat q^a,\hat p^a)$, defined as
\begin{align}
\begin{split}
 &\hat q^1     := \frac{\cosh  \hat  \eta^2  \cosh   \hat \eta^3 }{ \cosh\hat  \chi -1 }\,   \hat y^1  ,\qquad 
 \hat q^2  :=  \frac{\cosh    \hat\eta^3 }{  \cosh \hat \chi -1 } \,\hat y^2 ,\qquad 
\hat q^3     :=  \frac 1 {\cosh \hat \chi-1 }\, \hat y^3   ,\qquad 
  \hat p^a :=\hat  \eta^a  \,.
\end{split}
\label{bl}
\end{align}
Using the relations  (\ref{bi}) together with (\ref{bf}), one can indeed show that they fulfil   the canonical commutation relations
\begin{align}
\big[ \hat q^a,\hat q^b\big]= \big[\hat p^a,\hat p^b\big]= 0, \qquad   \big[ \hat q^a,\hat p^b\big]= \frac 1 {\kappa} \, \delta_{ab}  .
\label{bm}
\end{align}
Consequently, for $   {\mathcal{W} }_\kappa$  we recover  three copies of the usual Heisenberg--Weyl  algebra   of quantum mechanics where the deformation parameter $\kappa^{-1}$ replaces the Planck constant $\hbar$.  
Recall that the classical (Poisson) version of these coordinates was introduced in~\cite{BGH2019worldlinesplb}.

In addition, the complete $\kappa$-deformed  extended space $\widetilde   {\mathcal{W} }_\kappa$ can be expressed in terms of the Darboux operators $(\hat q^a,\hat p^a)$ by taking into account the following relation for $\hat y^0$ coming from (\ref{bi}):
\be
\hat y^0     = -  \left( \frac{\sinh   \hat p^1}{\cosh   \hat p^2  \cosh  \hat  p^3 }\, \hat q^1+ \frac{\cosh   \hat p^1 \sinh    \hat p^2 }{ \cosh   \hat p^3 }\, \hat q^2+   \cosh   \hat p^1 \cosh  \hat  p^2  \sinh   \hat p^3   \, \hat q^3   \right) .
\label{bn}
\ee
Therefore, the ``extended" phase space algebra associated to $\widetilde   {\mathcal{W} }_\kappa$ is given by~\eqref{bm} together with the commutators
\begin{align}
\begin{split}
&\big[\hat q^1, \hat y^0 \big]  =- \frac 1\kappa \left(  \frac{\cosh  \hat p^1 }{\cosh    \hat p^2 \cosh    \hat p^3 }\,\hat q^1+ \frac{\sinh    \hat p^1\sinh    \hat p^2 }{ \cosh    \hat p^3 }\, \hat q^2 +  {\sinh    \hat p^1 \cosh    \hat p^2  \sinh    \hat p^3}\, \hat q^3  \right) ,\\[2pt]
&\big[\hat q^2, \hat y^0 \big]  =  \frac 1\kappa \left(  \frac{\sinh    \hat p^1\tanh    \hat p^2 } {\cosh    \hat p^2 \cosh    \hat p^3}\, \hat q^1 - \frac{\cosh  \hat p^1\cosh    \hat p^2 }{\cosh    \hat p^3  } \,\hat q^2 -  { \cosh  \hat p^1  \sinh    \hat p^2  \sinh    \hat p^3}\, \hat q^3  \right) ,\\[2pt]
&\big[\hat q^3, \hat y^0 \big]  =  \frac 1\kappa \left(  \frac{\sinh    \hat p^1\tanh    \hat p^3 } {\cosh    \hat p^2 \cosh    \hat p^3}\, \hat q^1 + \frac{\cosh  \hat p^1\sinh    \hat p^2\tanh    \hat p^3 }{\cosh    \hat p^3  } \,\hat q^2 -  { \cosh  \hat p^1  \cosh    \hat p^2  \cosh  \hat p^3}\, \hat q^3  \right) ,
\end{split}
\label{bo}
\end{align}
and where  $[\hat p^a,\hat y^0   ]$ remain exactly the same as $[\hat \eta^a,\hat y^0   ]$ given in (\ref{bg}) since $\hat p^a\equiv \hat \eta^a$.


\section{Spaces of worldlines with $\kappa$-Galilei and $\kappa$-Carroll symmetries}
\label{s4}

The non-relativistic limit from the $\kappa$-Poincar\'e algebra to  the $\kappa$-Gailei one  can be  performed through  a Lie bialgebra contraction approach~\cite{BGHOS1995quasiorthogonal}   in which this limit is given by a contraction of the so-called fundamental  non-coboundary type. This contraction transforms the generators  according to the map (\ref{ab}), while the deformation parameter $\kappa$ remains unchanged. Taking the limit $c\to \infty$ the $r$-matrix (\ref{bc})  diverges but the cocommutators (and therefore the Hopf algebra structure) have a well-defined non-trivial limit (see~\cite{BGGH2020kappanewtoncarroll} for details). At the quantum group level, this contraction requires that  the quantum coordinates $\hat x^\mu,\hat y^\mu$ and $\hat \eta^a$ are mapped according to
\be
\hat x^a\to c\, \hat x^a,\qquad \hat y^a\to c\, \hat y^a,\qquad \hat \eta^a\to c\, \hat \eta^a,
\label{cca}
\ee
since these are coordinates associated to generators that are mapped in (\ref{ab}) with a $1/c$ factor; therefore, the inverse factor $c$ has to be introduced in~\eqref{cca} in order to keep the convergence of the corresponding exponentials~\eqref{bb}. 
Note that  $\hat x^0$ and $\hat y^0$ are not modified under the contraction map.

Concerning spacetime, it can be shown that the $\kappa$-Minkowski  commutators (\ref{a1}) are invariant under the contraction map (\ref{cca}). Therefore, the  $\kappa$-Galilei spacetime $\mathcal{G}_\kappa$ is algebraically identical to  $\kappa$-Minkowski  $\mathcal{M}_\kappa$. Nevertheless, the two models can be distinguished by looking at their associated spaces of worldlines. By applying the contraction map  (\ref{cca}) to  the $\kappa$-Poincar\'e space of worldlines $\mathcal{W}_\kappa$ (\ref{be}) and  taking the $c\to\infty$ limit, we find that the   $\kappa$-Galilei  space of worldlines is given by the following homogeneous quadratic algebra
\begin{equation}
\begin{split}
\big[\hat y^a, \hat y^b\big] &= \frac{1}{\kappa} \big(  \hat \eta^a\hat y^b   - { \hat \eta^b\hat y^a  } \big) ,\\
\big[\hat y^a, \hat \eta^b\big] &= \frac{1}{2\kappa} \, \delta_{ab}  \left(  ( \hat \eta^1)^2+  ( \hat \eta^2)^2+(  \hat \eta^3 )^2 \right), \\
\big[\hat \eta^a, \hat\eta^b\big] &= 0,\quad  \forall a,b . 
\end{split}
\label{ca}
\end{equation}
Notice that under  contraction
   \be
\lim_{c\to\infty}  c^2\left( \cosh(\hat\eta ^1/c )\cosh(\hat\eta ^2/c) \cosh(\hat\eta^3/c)-1\right)= \frac 12   \left(  ( \hat \eta^1)^2+  ( \hat \eta^2)^2+(  \hat \eta^3 )^2 \right) \equiv \frac 12\,  \hat { \boldsymbol{\eta}}^2 ,
 \label{cb}
 \ee
which is just the quantum analogue of (\ref{ak}).

It is worth mentioning that the algebra (\ref{ca}) strongly resembles  Snyder's model~\cite{Snyder1947}, provided that we   interpret  $(\hat y^a,\hat \eta^a)$ as  position- and momentum-type quantum variables. In this case, the first commutators can be written in terms of quantum angular momenta-type operators of the form
\be
\hat L_c=\epsilon_{abc}  \, { \hat y^a \hat \eta^b }\,,
 \label{cc1}
\ee 
such that 
\be
 \big[\hat y^a, \hat y^b\big]=-\frac 1 \kappa \, \epsilon_{abc}\, \hat L_c \, ,
 \label{cc2}
\ee 
while the commutators in the second line of  (\ref{ca})  can be written in terms of  the kinetic energy operator $\frac 12 \hat { \boldsymbol{\eta}}^2$. From this point of view
the   $\kappa$-Galilei  space of worldlines is algebraically similar to a hyperbolic Snyder--Euclidean model  (with negative curvature)~\cite{BGH2020snyder,BGH2020snyderPOS},  also known in the literature as a  ``non-relativistic" Snyder model~\cite{Mignemi:2011gr, Lu:2011it, Mignemi:2012gr,Ivetic:2015cwa}.

By following the same methodology we just outlined,  the non-relativistic limit of the $\kappa$-Poincar\'e extended  space of worldlines  $ \widetilde {\mathcal{W} }_\kappa$ (\ref{bg}) gives rise to additional relations of
 Lie-algebraic type
 \begin{equation}
 \big[\hat y^a, \hat y^0 \big] =  \frac 1{\kappa} \, \hat y^a ,\qquad   \big[\hat \eta^a, \hat y^0 \big] =  \frac 1{\kappa}\,  \hat \eta^a  .
\label{cd}
\end{equation}
Hence,   $(  \hat y^0, \hat \eta^a)$ close   a 4D linear subalgebra which is formally similar  to    the  $\kappa$-Galilei spacetime $\mathcal{G}_\kappa$~\eqref{a1}.  The self-consistency of the approach is verified by checking that $\mathcal{G}_\kappa$ is recovered  from the extended space $ \widetilde {\mathcal{W} }_\kappa$  by means of the following change of quantum variables
 \begin{equation}
 \hat x^0 = \hat y^0 ,\qquad   \hat x^a = \hat y^a  +  \hat \eta^a      \hat y^0   ,
\label{ce}
\end{equation}
which can also be obtained as a non-relativistic limit of  (\ref{bh}).

The differential realization of the seven generators  $(\hat y^\mu,\hat \eta^a)$ of  the $\kappa$-Galilei extended  space of worldlines $ \widetilde {\mathcal{W} }_\kappa$ as operators acting  on the space of functions $\Psi(\eta^1,\eta^2,\eta^3)$ reads
\begin{align}
\begin{split}
& \hat y^0  \Psi  = -\frac 1\kappa \left(     \eta^1 \,  \frac{\partial \Psi}{\partial \eta^1} +     \eta^2  \,\frac{\partial \Psi}{\partial \eta^2} +    \eta^3   \,  \frac{\partial \Psi}{\partial \eta^3}   \right) ,\\[2pt] 
& \hat y^a  \Psi  = \frac{1}{2\kappa} \,{ \boldsymbol{\eta}}^2\,   \frac{\partial \Psi}{\partial \eta^a}  ,\qquad  \hat \eta^a \Psi = \eta^a\Psi ,
\end{split}
\label{cf}
\end{align}
 where $\boldsymbol{\eta}^2$ is given by (\ref{ak}). Notice also that  (\ref{cf}) can  be obtained 
 through the non-relativistic limit  (\ref{cca}) from the representation (\ref{bi}). In analogy to what done in the $\kappa$-Poincar\'e case, from (\ref{cf}) we obtain the  representation of the spacetime coordinates of $\mathcal{G}_\kappa$  (\ref{a1}) on the  3-momentum space:
\be
\begin{split}
&\hat x^a  \Psi  = \frac 1\kappa \left(  \frac12\, \boldsymbol{\eta}^2\,  \frac{\partial \Psi}{\partial \eta^a} -\eta^a  \left(   \eta^1 \,  \frac{\partial \Psi}{\partial \eta^1} +     \eta^2  \,\frac{\partial \Psi}{\partial \eta^2} +    \eta^3   \,  \frac{\partial \Psi}{\partial \eta^3}  \right)  \right),\\
&\hat x^0  \Psi \equiv \hat y^0  \Psi\,.
\end{split}
\label{{cg}}
\ee

Quantum Darboux operators  $(\hat q^a,\hat p^a)$, which satisfy  (\ref{bm}), can be defined for the $\kappa$-Galilei  space of worldlines $   {\mathcal{W} }_\kappa$  as follows: 
\be
\hat q^a     := \frac{2 }{  \hat { \boldsymbol{\eta}}^2 }\,   \hat y^a  ,  \qquad  \hat p^a :=\hat  \eta^a  ,
\label{{ch}}
\ee
 In terms of these, the quantum coordinate $\hat y^0$ can be written as
\be
\hat y^0     = -  \left(   \hat p^1  \hat q^1+   \hat p^2  \, \hat q^2+      \hat p^3   \, \hat q^3   \right) .
\label{ci}
\ee
Therefore, the $\kappa$-Galilei extended space of worldlines $ {\mathcal{W} }_\kappa$ can be expressed via  the commutators 
\be
 \big[\hat q^a, \hat y^0 \big] =  - \frac 1{\kappa} \, \hat q^a ,\qquad   \big[\hat p^a, \hat y^0 \big] =  \frac 1{\kappa}\,  \hat p^a  \, ,
  \label{cj}
\ee 
together with  (\ref{bm}). These relations  are remarkably simpler than the corresponding $\kappa$-Minkowski relations (see, for instance,~\eqref{bo}).
 
The ultra-relativistic limit $c\to 0$ (or speed-time contraction~\cite{LevyLeblondCarroll,BLL})
from the $\kappa$-Poincar\'e algebra to the $\kappa$-Carroll algebra~\cite{CK4d,BGGH2020kappanewtoncarroll}  can be obtained as a Lie bialgebra contraction of coboundary type~\cite{BGHOS1995quasiorthogonal}. Besides mapping the  generators as described in (\ref{ad}), this contraction procedure requires that the deformation parameter is transformed as
\be
\kappa\to c\,\kappa .
\label{da}
\ee

In this way, the  classical $r$-matrix (\ref{bc}) is preserved  under contraction. For similar reasons as discussed in the $\kappa$-Galilei case, for the   quantum coordinates $\hat x^\mu$ and $(\hat y^\mu,\hat \eta^a)$ the contraction map is  given by
 \be
\hat x^0\to c^{-1} \hat x^0,\qquad \hat y^0\to  c^{-1} \hat y^0,\qquad \hat \eta^a\to  c^{-1}\hat \eta^a,
\label{db}
\ee
 preserving $\hat x^a$ and $\hat y^a$, and next applying the limit $c\to0$.   

Similarly to  the $\kappa$-Galilei case,  the $\kappa$-Minkowski commutators~\eqref{a1} are invariant under the contraction map~(\ref{da})--(\ref{db}). Therefore, the $\kappa$-Carroll spacetime  $\mathcal{C}_\kappa$ is algebraically identical to $\kappa$-Minkowski  $\mathcal{M}_\kappa$. 
Again, the two models can be distinguished by looking that the space of worldlines. From  (\ref{be}) one  finds that the $\kappa$-Carroll space of worldlines $\mathcal{W}_\kappa$  is commutative, while from (\ref{bg}) we deduce the   commutation relations  defining the $\kappa$-Carroll extended  space of worldlines $\widetilde{\mathcal{W}}_\kappa$, namely
\begin{align}
\begin{split}
& \big[\hat y^a, \hat y^b\big] =0, \qquad \big[\hat y^a, \hat \eta^b\big]=0,\qquad \big[\hat \eta^a, \hat\eta^b\big]= 0,\\[2pt] 
 &\big[\hat y^a, \hat y^0 \big] =  \frac 1{\kappa} \, \hat y^a ,\qquad   \big[\hat \eta^a, \hat y^0 \big] =  \frac 1{\kappa}\,  \hat \eta^a   .
\end{split}
\label{dc}
\end{align}
Equations~\eqref{dc} define the Lie algebra of a dilation operator $\hat y^0$ acting on the direct sum of two 3D Abelian algebras   $(  \hat y^a)$ and   $(  \hat \eta^a)$.  Consequently, Darboux operators cannot be constructed and, in fact, the quantum coordinates $\hat y^\mu$ and  $(  \hat y^0, \hat \eta^a)$  define two 4D Lie subalgebras which are isomorphic  to the $\kappa$-Carroll spacetime   $\mathcal{C}_\kappa$~\eqref{a1}. 
Finally, the map relating the  $\kappa$-Carroll spacetime  $\mathcal{C}_\kappa$~\eqref{a1} and the 
$\kappa$-Carroll extended space of worldlines $\mathcal{W}_\kappa$~\eqref{dc} reads
  \begin{equation}
\hat x^0 = \hat y^0 + \hat \eta^1  \hat y^1 + \hat \eta^2  \hat y^2 + \hat \eta^3  \hat y^3, \qquad \hat x^a = \hat y^a \, ,
\end{equation}
and can  be  straightforwardly  obtained by computing the ultra-relativistic limit~\eqref{db} of the Poincar\'e relations~\eqref{bh}.


\section{Concluding remarks}

In this paper we  constructed the spaces of time-like worldlines characterized by  quantum-deformed  $\kappa$-Galilei and $\kappa$-Carroll symmetries. We showed that these spaces allow to distinguish these kinematical models from the $\kappa$-Poincar\'e one, even though they share formally  the same $\kappa$-noncommutative spacetime~(\ref{a1}). 

By extending the spaces of worldlines to include the noncommutative time coordinate, we were able to define the map between the seven quantum coordinates on these spaces $(\hat y^\mu,\hat \eta^a)$ and the four quantum spacetime coordinates $\hat x^\mu$,   for the $\kappa$-Poincar\'e,   $\kappa$-Galilei and  $\kappa$-Carroll cases.
The possibility to find such a relation could be expected based on the classical intuition: points in spacetime can be defined by the intersection of worldlines, and, conversely, worldlines can be defined as trajectories in spacetime.
Thanks to this mapping, the extended  $\kappa$-Galilei space of worldlines defined by (\ref{ca}) and  (\ref{cd})  could  be used to describe non-relativistic fuzzy worldlines in a setting similar to~\cite{BGGM2021fuzzy}.

As a byproduct of this analysis, we found that the  $\kappa$-Poincar\'e and the $\kappa$-Galilei extended spaces of worldlines can be represented on a 3D  space defined by the classical variables $\eta^a$. These variables are associated to the worldlines rapidities and in this differential representation they play the role of momentum-type coordinates. The  space of rapidities  has a curved hyperbolic nature in the  $\kappa$-Poincar\'e case, while  it is  a flat Euclidean space in the $\kappa$-Galilei model. 
We would like to emphasize that this representation is quite different from the momentum space realizations which have been previously studied in the literature~\cite{KowalskiGlikman:2002ft,KowalskiGlikman:2003we,AmelinoCamelia:2011bm,Arzano:2014jua,Kowalski-Glikman:2013rxa,BGGHplb2017,BGGHplb2018}  in the context of deformed special relativity theories. We postpone to future work further investigations on the relation between these constructions.

As a final remark, we  stress that in this paper we  dealt with the usual ``time-like" $\kappa$-Poincar\'e deformation with classical $r$-matrix given by~\eqref{bc} along with its non-relativistic and ultra-relativistic limits. However, there also exist the  ``space-like" and ``light-like"  $\kappa$-deformations~\cite{BP2014extendedkappa}.  The   noncommutative spaces  of  all three types of $\kappa$-Poincar\'e geodesics have recently been  constructed in~\cite{BGH2022};  the approach here presented could be applied to the study of their Galilean and Carrollian limits, including the definition of the corresponding extended noncommutative spaces.

 \newpage
 

\section*{Acknowledgements }

\phantomsection
\addcontentsline{toc}{section}{Acknowledgements}

This work has been partially supported by Agencia Estatal de Investigaci\'on (Spain)  under grant  PID2019-106802GB-I00/AEI/10.13039/501100011033, by the Regional Government of Castilla y Le\'on (Junta de Castilla y Le\'on, Spain) and by the Spanish Ministry of Science and Innovation MICIN and the European Union NextGenerationEU/PRTR. The authors would like to acknowledge the contribution of the European Cooperation in Science and Technology COST Action CA18108.


\small



\begin{thebibliography}{99}


 
\phantomsection
\addcontentsline{toc}{section}{References}

 
 
\bibitem{Snyder1947}
  H.~S.~Snyder.
 Quantized space-time.
{\em  Phys.\ Rev.},  {71}:38--41, 1947.
   \href {http://doi.org/10.1103/PhysRev.71.38}
  {\path{doi:10.1103/PhysRev.71.38}}


   \bibitem{BGH2020snyder}
A.~Ballesteros, G.~Gubitosi, and F.~J. Herranz.
\newblock {Lorentzian Snyder spacetimes and their Galilei and Carroll limits
  from projective geometry}.
\newblock {\em Class. Quantum Gravity}, 37:195021,  2020.
\newblock \href {http://dx.doi.org/10.1088/1361-6382/aba668}
  {\path{doi:10.1088/1361-6382/aba668}}
  


 
   \bibitem{BGH2020snyderPOS}
A.~Ballesteros, G.~Gubitosi, and F.~J. Herranz.
\newblock {Generalized noncommutative Snyder spaces and projective geometry}.
\newblock {\em Proc. Science PoS (CORFU2019)}, 376:190,  2020.
\newblock \href {https://doi.org/10.22323/1.376.0190}
  {\path{doi:10.22323/1.376.0190}}


\bibitem{LRNT1991}
J.~Lukierski, H.~Ruegg, A.~Nowicki, and V.~N. Tolstoy.
\newblock {q-deformation of Poincar{\'{e}} algebra}.
\newblock {\em Phys. Lett. B}, 264:331--338, 1991.
\newblock \href {http://dx.doi.org/10.1016/0370-2693(91)90358-W}
  {\path{doi:10.1016/0370-2693(91)90358-W}}

\bibitem{GKMMK1992}
 S.~Giller, P.~Kosinski, M.~Majewski, P.~Maslanka, and J.~Kunz.
\newblock {More about the $q$-deformed Poincar{\'{e}} algebra}.
\newblock {\em Phys. Lett. B},  286:57--62, 1992.
\newblock \href {https://doi.org/10.1016/0370-2693(92)90158-Z}
  {\path{doi:10.1016/0370-2693(92)90158-Z}}

  \bibitem{LNR1992fieldtheory}
J.~Lukierski, A.~Nowicki, and H.~Ruegg.
\newblock {New quantum Poincar{\'{e}} algebra and $\kappa$-deformed field
  theory}.
\newblock {\em Phys. Lett. B}, 293:344--352, 1992.
\newblock \href {http://dx.doi.org/10.1016/0370-2693(92)90894-A}
  {\path{doi:10.1016/0370-2693(92)90894-A}}
  
    \bibitem{Maslanka1993}
P.~Maslanka.
\newblock {The $n$-dimensional $\kappa$-Poincar{\'{e}} algebra and group}.
\newblock {\em J. Phys. A: Math. Gen.}, 26:L1251--L1253, 1993.
\newblock \href {http://dx.doi.org/10.1088/0305-4470/26/24/001}
  {\path{doi:10.1088/0305-4470/26/24/001}}

\bibitem{MR1994}
S.~Majid and H.~Ruegg.
\newblock {Bicrossproduct structure of $\kappa$-Poincar{\'{e}} group and
  non-commutative geometry}.
\newblock {\em Phys. Lett. B}, 334:348--354, 1994.
\newblock \href {http://dx.doi.org/10.1016/0370-2693(94)90699-8}
  {\path{doi:10.1016/0370-2693(94)90699-8}}

\bibitem{Zakrzewski1994poincare}
S.~Zakrzewski.
\newblock {Quantum Poincar{\'{e}} group related to the $\kappa$-Poincar{\'{e}}
  algebra}.
\newblock {\em J. Phys. A: Math. Gen.}, 27:2075--2082, 1994.
\newblock \href {http://dx.doi.org/10.1088/0305-4470/27/6/030}
  {\path{doi:10.1088/0305-4470/27/6/030}}

\bibitem{BGGH2020kappanewtoncarroll}
A.~Ballesteros, G.~Gubitosi, I.~Gutierrez-Sagredo, and F.~J. Herranz.
\newblock {The $\kappa$-Newtonian and $\kappa$-Carrollian algebras and their
  noncommutative spacetimes}.
\newblock {\em Phys. Lett. B}, 805:135461, 2020.
\newblock 
  \href {http://dx.doi.org/10.1016/j.physletb.2020.135461}
  {\path{doi:10.1016/j.physletb.2020.135461}}
  
\bibitem{CK4d}
A.~Ballesteros,   F.~J. Herranz, M.~A. del Olmo, and M.~Santander.
\newblock {Four-dimensional quantum affine algebras and space-time $q$-symmetries}.
\newblock {\em J. Math. Phys.}, 35:4928--4940, 1994.
\newblock  \href{https://doi.org/10.1063/1.530823}{\path{doi:10.1063/1.530823}}

   \bibitem{Addazi:2021xuf}
A.~Addazi, J.~Alvarez-Muniz, R.~A.~Batista, \textit{et al.}
Quantum gravity phenomenology at the dawn of the multi-messenger era -- A review. {\em Prog. Part. Nucl. Phys.} 125:103948, 2022.
\href {https://doi.org/10.1016/j.ppnp.2022.103948}
  {\path{doi:10.1016/j.ppnp.2022.103948}}

\bibitem{BGH2019worldlinesplb}
A.~Ballesteros, I.~Gutierrez-Sagredo, and F.~J. Herranz.
\newblock {Noncommutative spaces of worldlines}.
\newblock {\em Phys. Lett. B}, 792:175--181, 2019.
\newblock  
 \href {http://dx.doi.org/10.1016/j.physletb.2019.03.029}
  {\path{doi:10.1016/j.physletb.2019.03.029}}

  
  \bibitem{BGGM2021fuzzy}
A. Ballesteros, G. Gubitosi, I. Gutierrez-Sagredo, and F. Mercati.
\newblock{Fuzzy worldlines with $\kappa$-Poincar{\'{e}} symmetries}.
\newblock {\em J. High Energ. Phys.}, 2021:80, 2021.
\href {https://doi.org/10.1007/JHEP12(2021)080}
 {\path{doi:10.1007/JHEP12(2021)080}}


\bibitem{BGHOS1995quasiorthogonal}
A.~Ballesteros, N.~A. Gromov, F.~J. Herranz, M.~A. del Olmo, and M.~Santander.
\newblock {Lie bialgebra contractions and quantum deformations of
  quasi-orthogonal algebras}.
\newblock {\em J. Math. Phys.}  36:5916--5937, 1995.
\newblock  \href {http://dx.doi.org/10.1063/1.531368}
  {\path{doi:10.1063/1.531368}}



 \bibitem{Inonu:1953sp}
  E.~In\"on\"u and E.~P.~Wigner.
 \newblock  On the Contraction of groups and their representations.
 \newblock{\em Proc.\ Nat.\ Acad.\ Sci.},  {39}:510--524, 1953.
\newblock \href{https://doi.org/10.1073/pnas.39.6.510}{\path{doi:10.1073/pnas.39.6.510}}

\bibitem{BLL}
H.~Bacry and J.~M.~L\'evy-Leblond.
\newblock{Possible Kinematics}.
\newblock{\em J. Math. Phys.}, {9}:1605--1614, 1968.
\newblock \href {http://dx.doi.org/10.1063/1.1664490}
  {\path{doi:10.1063/1.1664490}}


\bibitem{LevyLeblondCarroll} 
J.~M.~L\'evy-Leblond.
\newblock Une nouvelle limite non-relativiste du group de Poincar\'e.
\newblock {\em  Ann. Inst. H.~Poincar\'e}, 3:1--12,  1965. 
\url{http://www.numdam.org/item/AIHPA_1965__3_1_1_0/}

   \bibitem{conf}
 F.~J.~Herranz  and  M.~Santander.
\newblock Conformal symmetries of spacetimes.
\newblock{\em  J. Phys. A: Math. Gen.}, 35:6601--6618, 2002.
\newblock \href {http://dx.doi.org/10.1088/0305-4470/35/31/306}
  {\path{doi:10.1088/0305-4470/35/31/306}}
  
    \bibitem{Duval2014}
C.~Duval, G.~W. Gibbons, P.~A. Horvathy, and P.~M. Zhang.
\newblock {Carroll versus Newton and Galilei: two dual non-Einsteinian concepts
  of time}.
\newblock {\em Class. Quantum Gravity}, 31:085016,   2014.
\newblock \href {http://dx.doi.org/10.1088/0264-9381/31/8/085016}
  {\path{doi:10.1088/0264-9381/31/8/085016}}



\bibitem{HS1997phasespaces}
F.~J. Herranz and M.~Santander.
\newblock {Homogeneous phase spaces: the Cayley--Klein framework}.
\newblock In J.~F. Cari{\~{n}}ena, E.~Martinez, and M.~F. Ra{\~{n}}ada (Eds), {\em Geometr\'ia y F\'isica. Memorias la Real Acad. Ciencias}, vol.~XXXII,
  pp.~59--84, Madrid, 1998.
\newblock \href {http://arxiv.org/abs/physics/9702030}
  {\path{arXiv:physics/9702030}}


\bibitem{ChariPressley1994}
V.~Chari and A.~Pressley.
\newblock {\em {A guide to quantum groups}}.
\newblock Cambridge University Press, Cambridge, 1994.

  

\bibitem{Mignemi:2011gr}
  S.~Mignemi.
  Classical and quantum mechanics of the nonrelativistic Snyder model.
 {\em Phys.\ Rev.\ D},  {84}:025021,  2011.
   \href {http://doi.org/10.1103/PhysRevD.84.025021}
  {\path{doi:10.1103/PhysRevD.84.025021}}
  
  
    \bibitem{Lu:2011it}
  L.~Lu and A.~Stern.
  Snyder space revisited.
  {\em Nucl.\ Phys.\ B}, {854}:894--912, 2012.
   \href {http://doi.org/10.1016/j.nuclphysb.2011.09.022}
  {\path{doi:10.1016/j.nuclphysb.2011.09.022}}
 
  \bibitem{Mignemi:2012gr}
  S.~Mignemi.
Classical and quantum mechanics of the nonrelativistic Snyder model in curved space.
 {\em Class. Quant. Grav.},  {29}:215019,  2012.
  \href {https://doi.org/10.1088/0264-9381/29/21/215019}
  {\path{doi:10.1088/0264-9381/29/21/215019}}

 
  
\bibitem{Ivetic:2015cwa}
  B.~Iveti\'c, S.~Mignemi, and A.~Samsarov.
 Spectrum of the hydrogen atom in Snyder space in a semiclassical approximation.
{\em  Phys.\ Rev.\ A}, {93}:032109,  2016.
   \href {http://doi.org/10.1103/PhysRevA.93.032109}
  {\path{doi:10.1103/PhysRevA.93.032109}}

  
\bibitem{KowalskiGlikman:2002ft}
  J.~Kowalski-Glikman.
De sitter space as an arena for doubly special relativity.
 {\em  Phys.\ Lett.\ B},  {547}:291--296,  2002.
   \href {http://doi.org/10.1016/S0370-2693(02)02762-4}
  {\path{doi:10.1016/S0370-2693(02)02762-4}}


\bibitem{KowalskiGlikman:2003we}
  J.~Kowalski-Glikman and S.~Nowak,
 Doubly special relativity and de Sitter space.
{\em  Class.\ Quant.\ Grav.},   20:4799--4816, 2003.
   \href {http://doi.org/10.1088/0264-9381/20/22/006}
  {\path{doi:10.1088/0264-9381/20/22/006}}


 
\bibitem{AmelinoCamelia:2011bm}
  G.~Amelino-Camelia, L.~Freidel, J.~Kowalski-Glikman, and L.~Smolin,
Principle of relative locality.
{\em   Phys.\ Rev.\ D},  {84}:084010,  2011.
   \href {http://doi.org/10.1103/PhysRevD.84.084010}
  {\path{doi:10.1103/PhysRevD.84.084010}}

  

\bibitem{Arzano:2014jua}
  M.~Arzano, G.~Gubitosi, J.~Magueijo,  and G.~Amelino-Camelia.
Anti-de Sitter momentum space.
{\em   Phys.\ Rev.\ D},  {92}:024028,  2015.
   \href {http://doi.org/10.1103/PhysRevD.92.024028}
  {\path{doi:10.1103/PhysRevD.92.024028}}


   
    \bibitem{Kowalski-Glikman:2013rxa}
  J.~Kowalski-Glikman.
Living in Curved Momentum Space.
{\em   Int.\ J.\ Mod.\ Phys.\ A},  {28}:1330014,  2013.
  \href {http://doi.org/10.1142/S0217751X13300147}
  {\path{doi:10.1142/S0217751X13300147}}
 


\bibitem{BGGHplb2017}  
A.~Ballesteros, G.~Gubitosi, I.~Gutierrez-Sagredo, and  F.~J.~Herranz.
Curved momentum spaces from quantum groups with cosmological constant.
{\em Phys.\ Lett.\ B},  {773}:47--53, 2017.
   \href {http://doi.org/10.1016/j.physletb.2017.08.008}
  {\path{doi:10.1016/j.physletb.2017.08.008}}

   
\bibitem{BGGHplb2018}  A.~Ballesteros, G.~Gubitosi, I.~Gutierrez-Sagredo, and  F.~J.~Herranz.
Curved momentum spaces from quantum (Anti-)de Sitter groups in (3+1) dimensions.
{\em Phys. Rev. D},  97:106024, 2018.
   \href {https://doi.org/10.1103/PhysRevD.97.106024}
  {\path{doi:10.1103/PhysRevD.97.106024}}

\bibitem{BP2014extendedkappa}
A.~Borowiec and A.~Pachol.
\newblock {$\kappa$-Deformations and extended $\kappa$-Minkowski spacetimes}.
\newblock {\em Symmetry, Integr. Geom. Methods Appl.}, 10:107, 2014.
\newblock \href {http://dx.doi.org/10.3842/SIGMA.2014.107}
  {\path{doi:10.3842/SIGMA.2014.107}}

\bibitem{BGH2022}  A.~Ballesteros,  I.~Gutierrez-Sagredo, and  F.~J.~Herranz.
All noncommutative spaces of $\kappa$-Poincar\'e geodesics.
 \newblock{\em  J. Phys. A: Math. Theor.}, 55:435205, 2002.
\newblock \href {http://dx.doi.org/10.1088/1751-8121/ac99af}
  {\path{doi:10.1088/1751-8121/ac99af}}


 


      
\end{thebibliography}
\end{document}